# Central recirculation zone in a V-shaped premixed swirling flame

Qiuxiao Wang, Yongzhi Ren, Mingming Gu, Bowen Yu, Xiaoxing Feng, Fei Qi, Xi Xia*

School of Mechanical Engineering, Shanghai Jiao Tong University,

Shanghai 200240, P. R. China

**Abstract**

This paper presents an experimental study on the emergence of the central recirculation zone (CRZ) in a V-shaped premixed swirling flame, using simultaneous measurement of particle image velocimetry (PIV) and CH* chemiluminescence. The results show that either increasing the Reynolds number ($Re$) or decreasing the equivalence ratio ($\Phi$) would facilitate the emergence of CRZ. Further analysis demonstrates that the CRZ characteristics and its emergence are strongly influenced by the inner shear layer (ISL) surrounding the CRZ, while the swirl intensity remains unchanged. Dimensional analysis is performed to understand the underlying mechanism, suggesting the CRZ emergence is controlled by a non-dimensional parameter, $Re_s = |\gamma|_{max} D / \nu_s$, defined based on the maximum ISL intensity ($|\gamma|_{max}$), the exit diameter ($D$), and the kinematic viscosity ($\nu_s$) of the burnt gas. By estimating the temperature and viscosity with a simple heat-loss model, we show in the $|\gamma|_{max} D - \nu_s$ regime diagram that the cases with and without CRZ are separated by a single boundary line, corresponding to a critical $Re_s$ of about 424. This verifies the applicability of the proposed $Re_s$ criterion to lean-premixed V-shaped swirling flames under various conditions. Unlike most previous works that attribute the CRZ of swirling flames to vortex breakdown, the present work reveals the non-negligible effect of the ISL, especially the CRZ suppression when the ISL is weakened by flame heating.

---

* Corresponding author
E-mail address: xiaxiss@sjtu.edu.cn (Xi Xia)

# I. INTRODUCTION

Swirl combustion is prevalently adopted in power and propulsion systems, such as gas turbines and aero-engines [1-3]. The essence of applying swirl in combustion is to create a central recirculation zone (CRZ) [4], where the temperature is sufficiently high to serve as a continuous ignition source and the flow speed is sufficiently low to be matched with the flame speed, thereby promoting flame anchoring and stabilization [5]. Moreover, the reactants and active intermediates can be recycled in the CRZ, resulting in a longer residence time and thereby contributing to an improved combustion efficiency [6].

Essentially, the swirl-induced CRZ is a fluid dynamical phenomenon resulting from vortex breakdown [5,7]. It occurs as a strong swirling vortex tube suddenly expands and ruptures into a cone or bubble, inside which flow reverses to form a recirculation zone [8]. Benjamin [7] believed that vortex breakdown is related to inertial wave propagation, which triggers an abrupt change between two conjugate states of a cylindrical vortex. Gartshore [9] and Grabowski and Berger [10] proposed that the occurrence of vortex breakdown is due to the diffusion or convection of vorticity away from the vortex core. Leibovich et al. [11] and Lessen et al. [12] concluded that vortex breakdown could be interpreted as a consequence of hydrodynamic instability. Other understandings were more or less based on some combinations of the above theories [13-15]. The basic characteristics of CRZ in swirling flows have also been extensively studied [5,16,17]. Syred and Beer [5] experimentally found that the width of the CRZ increases with the swirl number, while the length of the CRZ first increases and then decreases with the swirl number. They also found that a divergent exit nozzle can create a larger CRZ and substantially reduce the threshold swirl number to trigger the onset of CRZ. Wang *et al.* [17] reported that the width and length both increase with the Reynold number ($Re$). Nogenmyr *et al.* [16] concluded that the existence of confinement could significantly enlarge the CRZ. Mukherjee *et al.* [18] numerically found that decreasing the radius of the center body can promote the onset of CRZ precession.

In swirling flames, the formation mechanism and characteristics of CRZ should be largely consistent with those for the non-reacting flow case. However, the behaviors of CRZ can be substantially affected by combustion through two main mechanisms. One is associated with the heat release process, which causes thermal expansion of burned gases and results in significantly accelerated flow downstream of the flame front; the other is related to the aftermath of heat release, i.e., the increased gas temperature, which influences the fluid dynamics through changed fluid properties, such as density and viscosity. For example, Gouldin *et al.* [19] experimentally studied the flow characteristics in a swirl combustor and found that combustion can promote the onset of vortex breakdown. They explained that combustion affects the flow substantially to cause a transition from supercritical to transcritical flow, thereby allowing the formation of a CRZ. Choi *et al.* [20] numerically found that the critical swirl number for the emergence of CRZ first decreases and then increases as the exothermicity is increased. They attributed this behavior to the nonlinear interaction between the advection of azimuthal vorticity and the baroclinic effects resulting from the coupling between the density and temperature gradients. Umeh *et al.* [21] experimentally studied the effects of equivalence ratio ($\Phi$) on CRZ in swirling flames and found that the CRZ size reduces and its position moves downstream with increasing $\Phi$. Their theoretical discussion attributes the effects of reaction on the CRZ to the complicated competition between effects that produce azimuthal vorticity along the vortex core, including the tilting and stretching of vorticity by velocity gradients and by dilatation, and the various baroclinic effects. Wang *et al.* [22,23] recently found that the length of CRZ in single swirling flames is shorter compared with the cold flow, but it could increase in the case of stratified swirling flames. They also observed spiral-shaped CRZ in stratified swirling flames and found that the outer main flame has a marginal influence on the CRZ compared with the inner pilot flame [24].

Albeit numerous research has been conducted, an in-depth and consistent understanding of the formation and characteristics of CRZ in basic swirling flames remains out of reach. This is probably because most relevant studies were conducted based on engineering-derived combustors [21,25], which usually involve complex structures for the inner duct and nozzle. For example, the swirler could induce significant wake vortices, which turn the flow into turbulence even at a small Reynolds number; a divergent nozzle or a central bluff body could promote the formation of CRZ. These additional factors make it difficult to set up clean boundary conditions for a scientific problem, let alone extract the core mechanisms or draw general conclusions.

Given this background, this study is motivated by a simple question: what determines the emergence of CRZ in a general swirling flame? To this end, we first develop a novel swirl burner to generate a basic swirling flame, which is free from the influences of hetero nozzle shape, central bluff body, or turbulent wake after the swirler. Then, the CRZ characteristics of lean-premixed methane-air swirling flames are experimentally investigated for wide ranges of Reynolds number and equivalence ratio, based on simultaneous measurement of particle image velocimetry (PIV) and flame chemiluminescence. The results enable us to identify a threshold condition based on the inner shear layer intensity, which could shed light on the general CRZ formation mechanism in the lean combustion regime.

## II. EXPERIMENTAL METHOD

The experimental setup consists of a novel single-nozzle open-head swirl burner of methane/air premixture and a simultaneous PIV/CH* chemiluminescence diagnostic system. As shown in Fig. 1, the burner contains a radial swirler of cambered-NACA-airfoil arrays to induce a swirling flow with minimum wake. The nozzle exit has a diameter of $D = 20$ mm, and features a convergent design with a streamlined center cone to suppress the upstream flow separation. For further suppression of upstream turbulence, the premixed gas was decelerated via a divergent annular section, ensuring that the flow impinges on the swirler vanes in a low-speed laminar regime. Both fuel and air streams were regulated by mass flow controllers (SevenStar series) at 1atm and 298K. By setting the fuel and air flowrates, the operation conditions for the present experiment were adjusted in terms of the Reynolds number ($Re$, range: 2000~7000) and the equivalence ratio ($\Phi$, range: 0.5~1.0). Here, $Re = UD/\nu$ is defined based on the bulk velocity ($U$) at the nozzle exit, the exit diameter ($D$), and the kinematic viscosity ($\nu$) of the fresh mixture; $\Phi = S_{toic} m_f / m_a$ is calculated based on the stoichiometric air-to-fuel mass ratio ($S_{toic} = 17.11$ for methane/air combustion), and the mass flow rates of fuel ($m_f$) and air ($m_a$). The maximum uncertainties in the measurements of $Re$ and $\Phi$ are 2.74% and 1.61%, respectively.

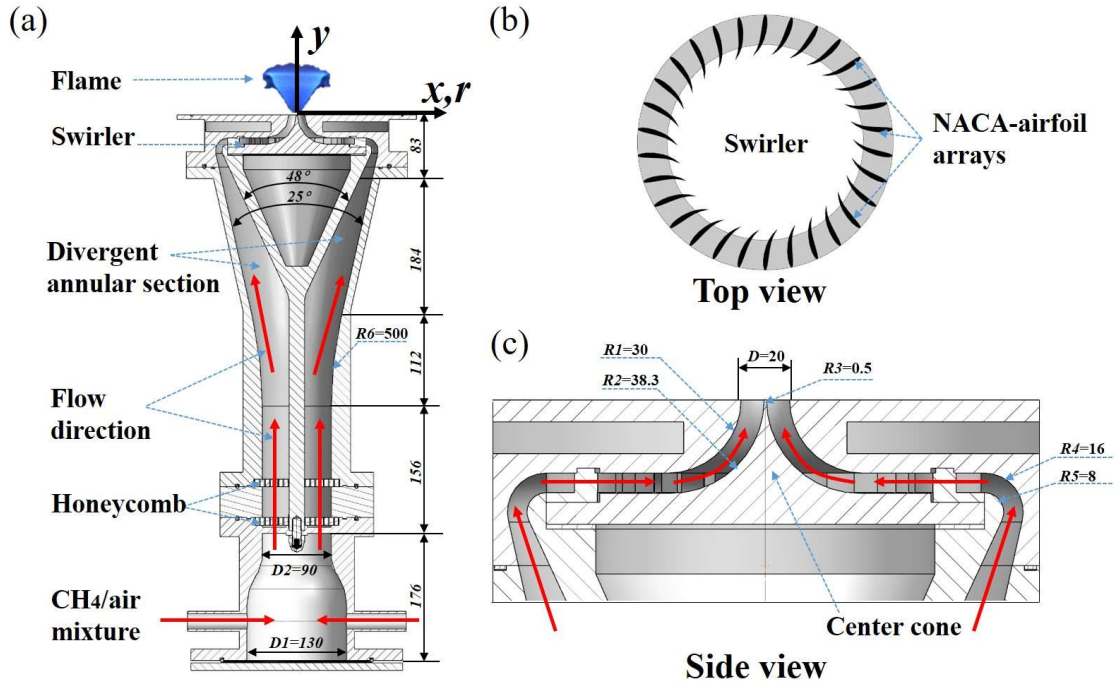

FIG. 1. Schematics of the swirl burner.

Fig. 2 presents the laser diagnostic system, the core of which is a dual-cavity Nd:YLF laser (Beamtech Vlite-Hi-527), which outputs green light of 527 nm at 3.1 mJ/pulse and a repetition rate up to 10 kHz. Two high-speed CMOS cameras (Phantom VEO1310L) with a maximum frame rate of 10.8 kHz at full resolution (1280×960 pixels) were used for PIV and chemiluminescence imaging, respectively. For PIV measurement, aluminum oxide particles of ~1 μm diameter were seeded in the air stream as flow tracers (Stokes number <0.023). Particle images were recorded at 10,000 fps by the PIV camera with a Nikkor lens (Nikkor 50 mm f/1.4G) and a short bandpass filter (Edmund Optics 527/20 nm). The field-of-view (FOV) was about 80 mm × 60 mm with a resolution of ~62 μm /pixel. For the main study, the laser sheet was aligned with the central verticle plane of the swirling flame. To evaluate the swirl number, the laser sheet was expanded horizontally to measure the transverse velocity field at the cross-section 2 mm above the nozzle exit. The velocity vectors were computed using a multi-scale cross-correlation PIV algorithm (LaVision Davis 8.0), with a final interrogation window size of 24 × 24 pixels and 50% overlap, resulting in a flow-field resolution of 0.744 mm/point; the maximum uncertainties in the velocity and vorticity measurements are estimated to be 1.84% and 3.37%, respectively. To simultaneously capture the flame structure, CH* chemiluminescence was recorded by the CH* camera attached with an intensifier (Lambert HiCATT) and a short bandpass filter (Edmund Optics, 430/10 nm). For each run, the burner operated for ~6 minutes after ignition so that the flame achieves thermal equilibrium and a fully-developed state, and then 5800 snapshots were captured for PIV and CH* imaging, respectively.

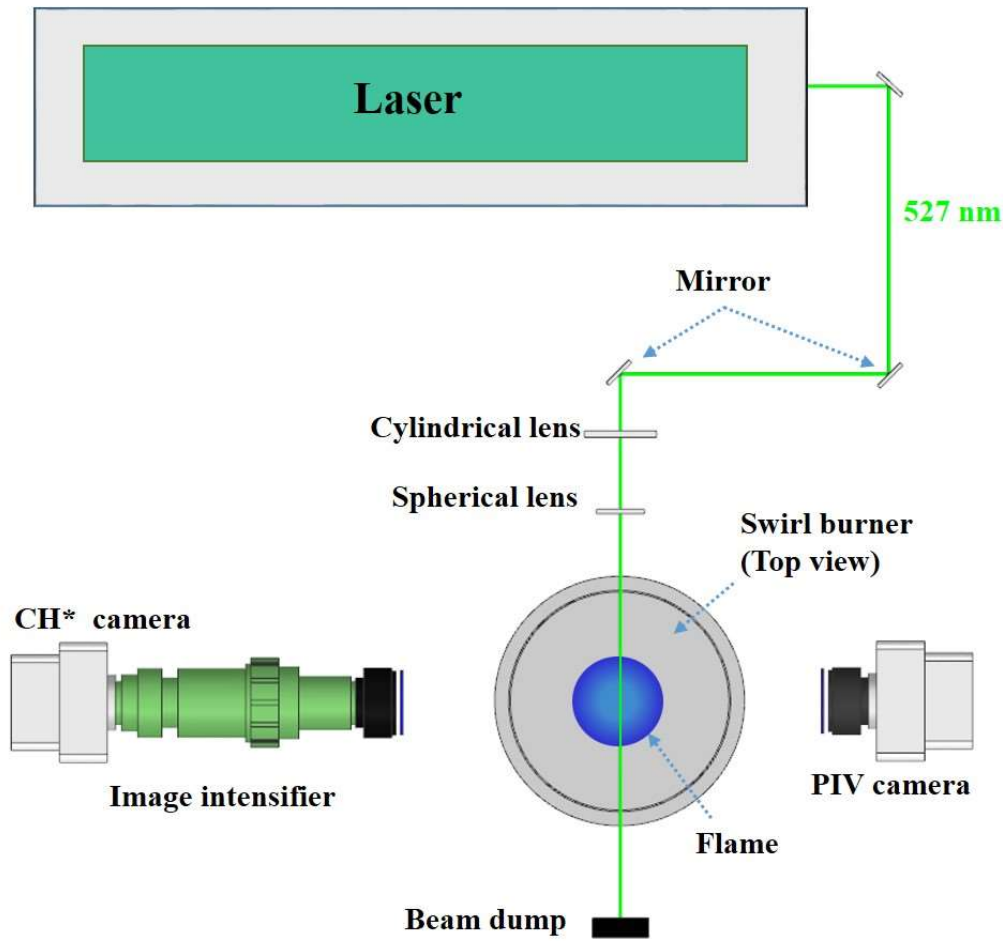

FIG. 2. Schematic of the laser diagnostic system.

## III. PARAMETRIC STUDY OF CRZ EMERGENCE

### A. Phenomenological description and regime nomogram

We first seek a phenomenological interpretation of the CRZ in swirling flames, based on the simultaneous PIV/CH* chemiluminescence system measurement. The flame and flow evolutions of two representative cases are presented in Fig. 3(a) and 3(b), corresponding to situations with and without CRZ, respectively. In the left half of each subplot, Abel inversion [26] of the CH* chemiluminescence signal is first performed to expose the flame front; then, the flame front is extracted and overlapped with the vorticity field and streamlines in the right-half subplot. For the lean-premixed combustion concerned in the present work, the swirling flames feature a V-shaped flame front, a weak inner shear layer (ISL), and a strong outer shear layer (OSL). The ISL decays, and its vorticity diffuses as it evolves downstream, whereas the OSL rolls up into vortex rings owing to the Kelvin–Helmholtz instability. The V-shaped flame front is anchored at the inner root of the ISL and then extends across the ISL to interact with the OSL in the far field. The coupling between the flame front roll-up and the outer vortex ring (OVR) movement confirms our previous finding [22,23] that the OVR dominates the flame dynamics and heat release rate fluctuation of a V-shaped flame. Unlike a fully turbulent flame, whose flame front is irregular and even fragmented owing to the turbulent eddies of different sizes, the flame fronts of the V-shaped swirling flames in Fig. 3 are relatively smooth, and the flame roll-up behaviors render periodic dynamics owing to the coherent OVRs developed from the OSL; these indicate that the present flow and flame starts from laminar in the near field and gradually evolves to be turbulent in the downstream. A notable difference between the two cases lies in the presence of a CRZ in Fig. 3(a), which is enclosed by the ISL; however, no CRZ can be found in Fig. 3(b). Meanwhile, the flame in Fig. 3(a) lasts up to more than 40 mm downstream, whereas the flame height in Fig. 3(b) is around 30mm. Apparently, different V-shaped flames could have different CRZ characteristics, which intrigues us to understand the formation mechanism of CRZ, especially, what determines its emergence.

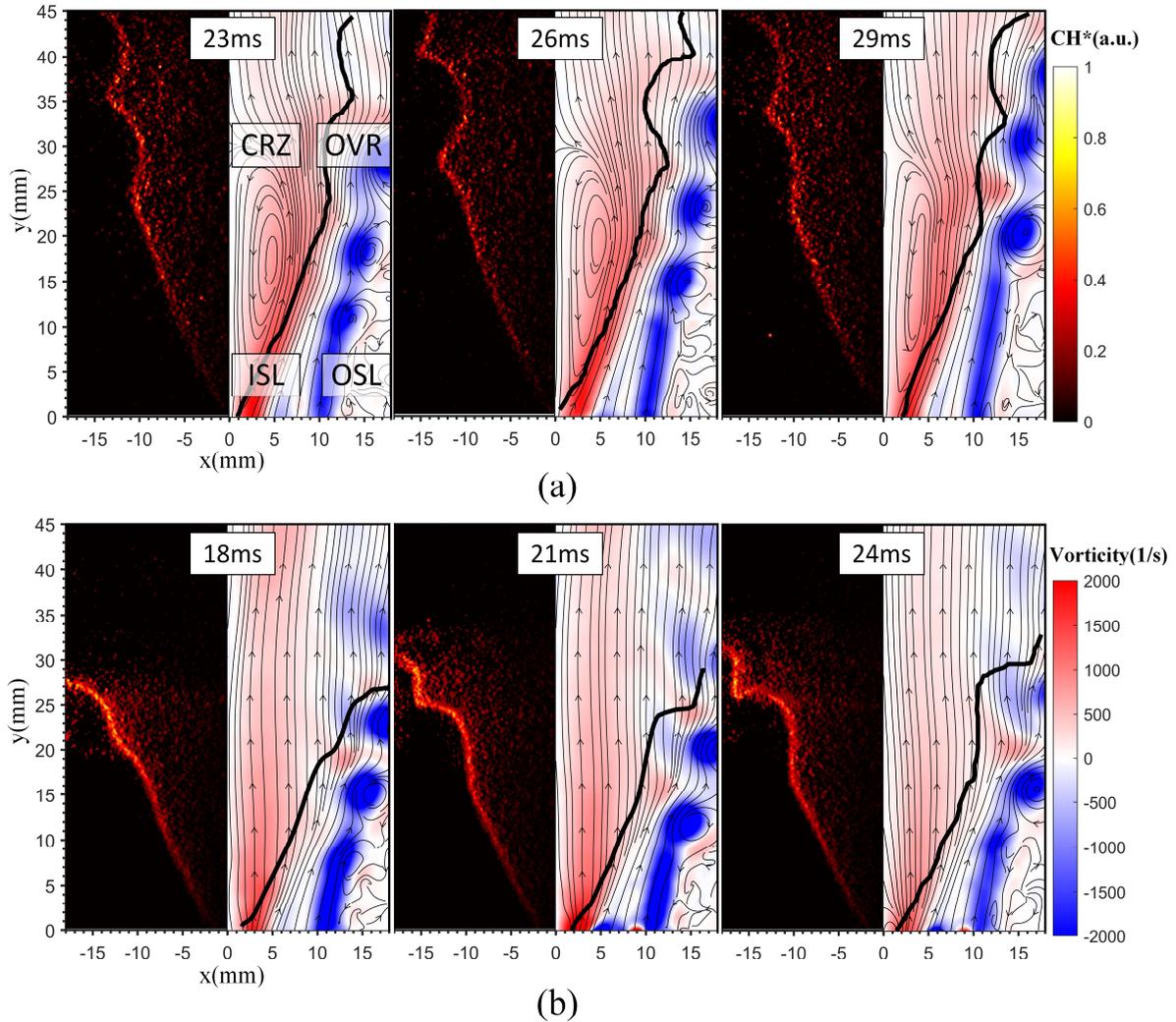

FIG. 3. The evolutions of the flame front (left) and flow field (right) for two representative V-shaped flames, (a) with CRZ at $Re = 3000$ and $\Phi = 0.56$, and (b) without CRZ at $Re = 4000$ and $\Phi = 0.8$. In each snapshot, the left-half image is the Abel inversion of the CH* chemiluminescence field; the right-half plot shows the vorticity field overlapped by the streamlines, where the black line marks the flame front extracted based on the ridge of the CH* intensity field.

In this work, we investigate the CRZ emergence and its influencing factors with different swirling flames obtained in a wide parameter space of $Re$ and ($Re$: 2000~7000, $\Phi$: 0.5~1.0), which roughly covers the lean-premixed V-shaped flame in the laminar-to-turbulent transition regime. The parameters outside the above range were not studied as the flame could take different modes. For example, the flame could turn to M-shaped when $\Phi$ is larger than 1, and become lifted when $Re$ is larger than 7000. The resulting $Re - \Phi$ regime nomogram is shown in Fig. 4, where the black dashed line denotes the critical boundary line between the regimes with and without CRZ. We can observe that either increasing $Re$ or decreasing $\Phi$ could promote the emergence of the CRZ, and the critical $Re$ increases with $\Phi$. Moreover, the limiting behaviors of the critical line at both ends seem to indicate that there could exist a low-$\Phi$ threshold and a high-$Re$ threshold, beyond which the CRZ would always appear. Given the relevance of $Re$ and $\Phi$ to the fluid dynamical and chemical effects, respectively, the formation of the CRZ may be viewed as a competition between the positive effect of jet momentum and the negative effect of the reaction rate; the detailed mechanism will be explored in the following.

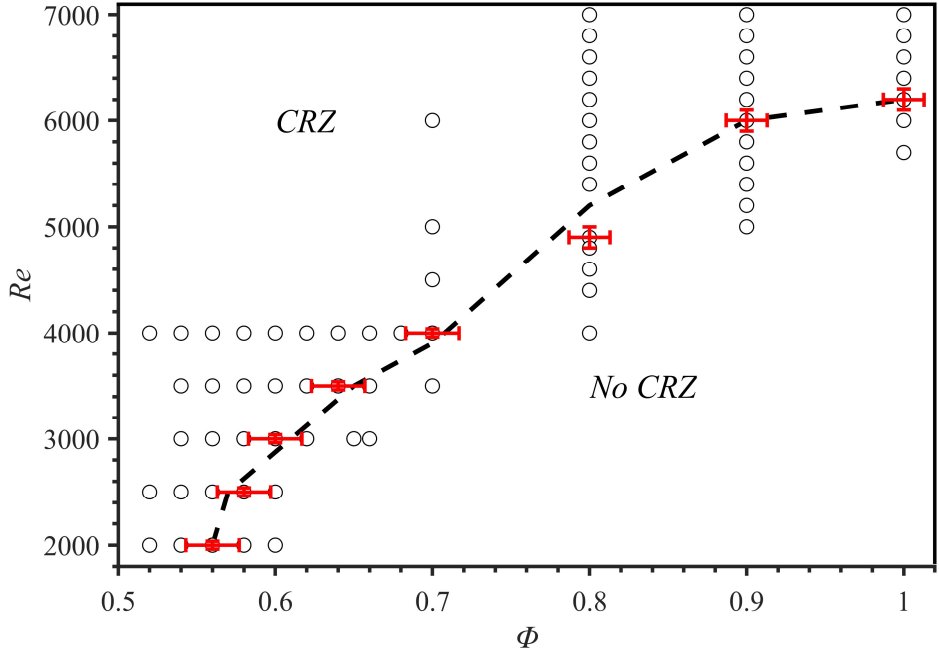

FIG. 4. Regime nomogram for CRZ emergence in the $Re - \Phi$ parameter space, based on 69 sets of experimental data. The dashed line denotes an approximate boundary between the flame regimes with and without the CRZ. The error bars are plotted for the data points in the vicinity of the boundary.

**B. Effects of key parameters on CRZ emergence**

We now separately study the effects of $Re$ or $\Phi$ on the characteristics of CRZ. Test runs were first carried out for various equivalence ratios at fixed $Re$ of 2500, 3000, and 3500. For quantitative estimation of the CRZ characteristics, the flow field and the flame front were analyzed in a time-averaged manner. In Fig. 5(a), we can readily observe that for a fixed $Re$ of 2500 the size of the CRZ first decreases with increasing $\Phi$ and then disappears at a critical equivalence ratio of $\Phi = 0.59$. Meanwhile, the flame height also decreases with increasing $\Phi$, owing to the larger reaction rate, which causes a faster reactant consumption than transportation. Similar trends of CRZ and flame shape can also be observed for cases of $Re=3000$ and $Re=3500$, with the critical equivalence ratios being 0.65 and 0.68, respectively. To further investigate how $Re$ influences the CRZ, experiments were performed for various *Re* at fixed equivalence ratios of 0.7, 0.8, and 1.0. It can be observed from Fig. 6 that for a fixed equivalence ratio the CRZ appears and then grows in size with increasing *Re*; and the flame grows taller with *Re* owing to the higher streamwise jet velocity, which brings the reactants farther downstream within a certain reaction period. In Figs. 5 and 6, we also calculate the flame base angles using a similar method to O'Connor and Lieuwen [27], resulting in a relatively constant angle of around 27° for all cases.

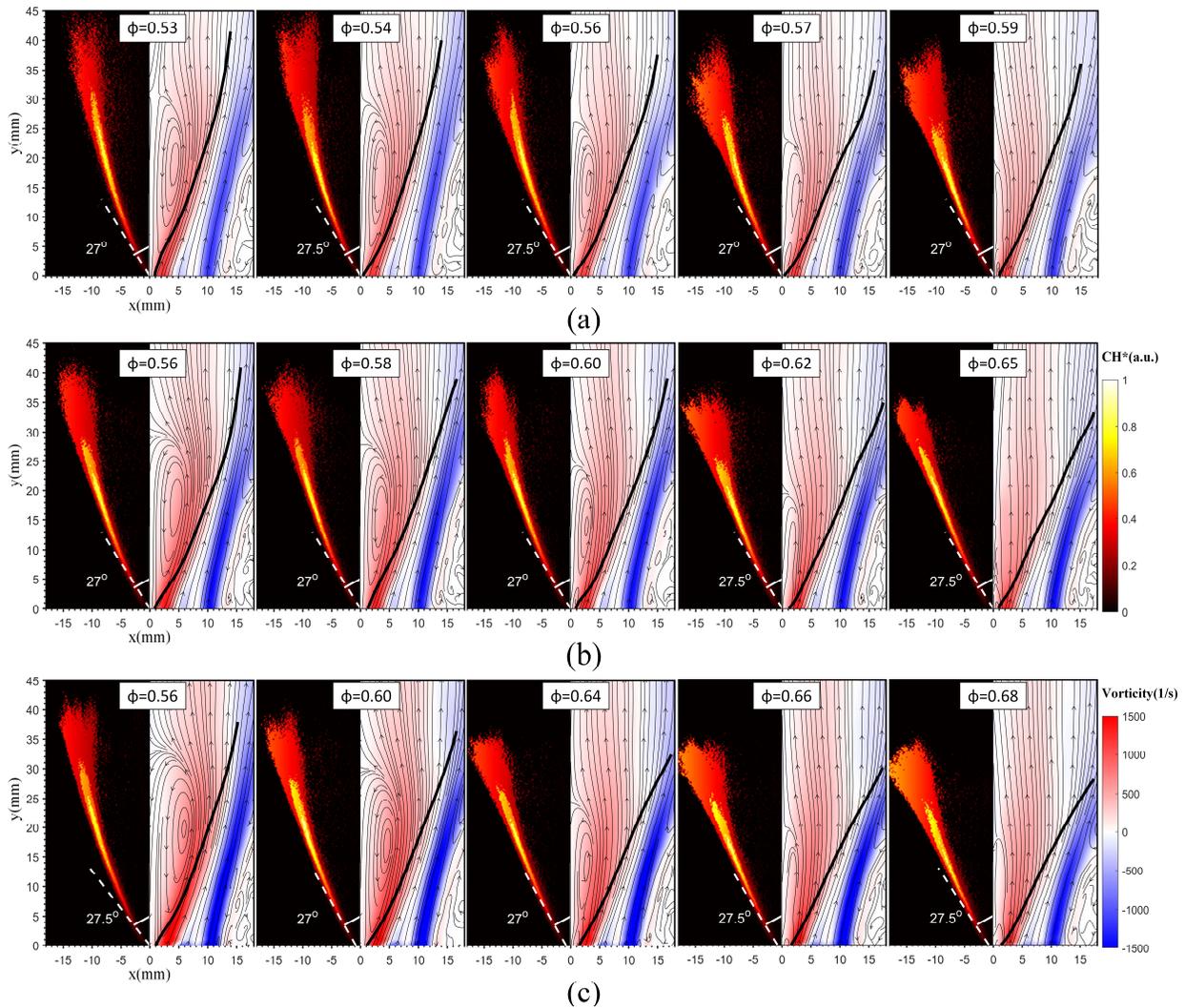

FIG.5. Comparisons of the time-averaged flame front (left) and flow field (right) for various equivalence ratios at (a) *Re* = 2500, (b) *Re* = 3000, and (c) R*e* = 3500.

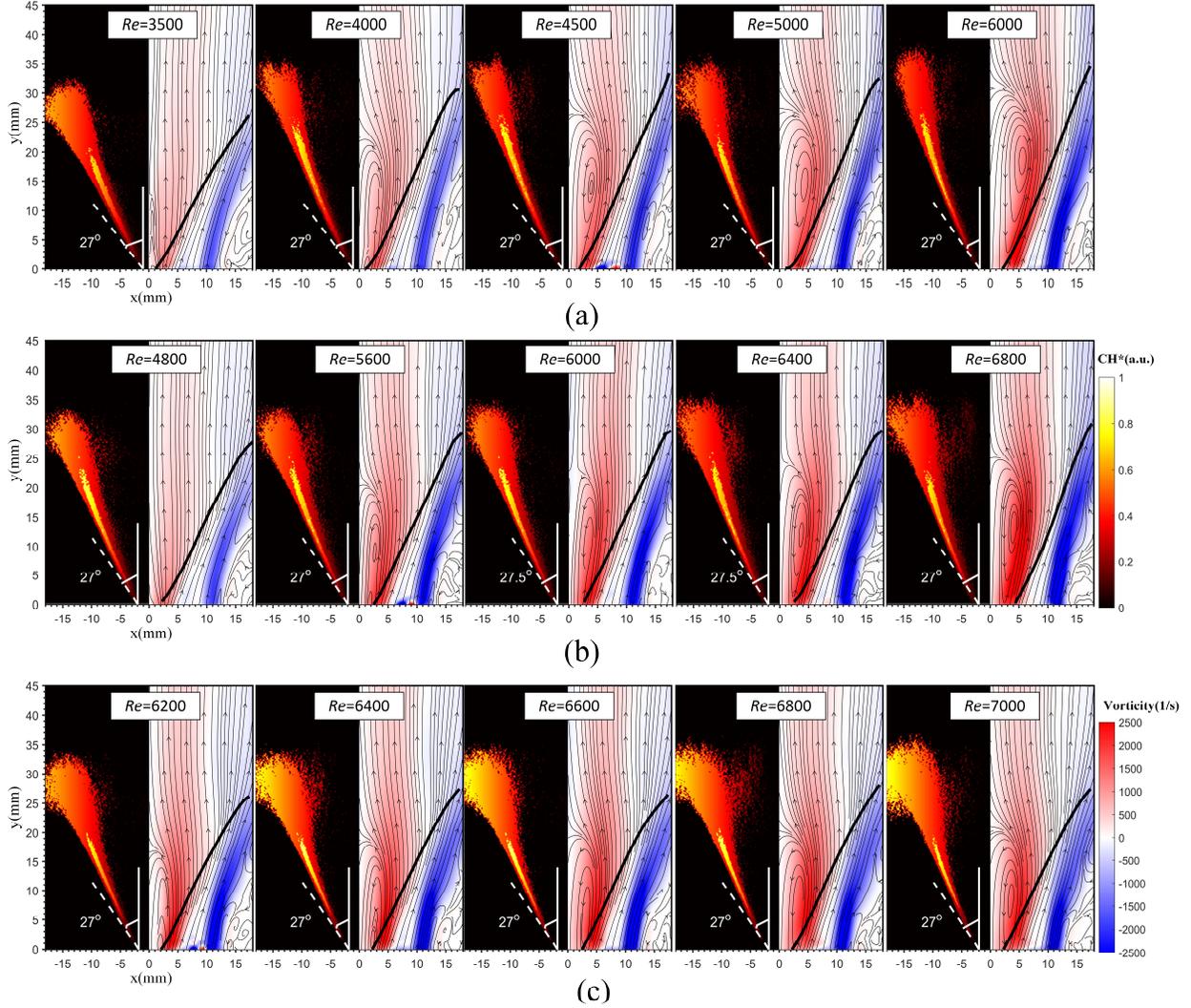

FIG.6 Comparisons of the time-averaged flame front (left) and flow field (right) for various Reynolds numbers at (a) $\Phi = 0.7$, (b) $\Phi = 0.8$, and (c) $\Phi = 1.0$.

To quantify the above influences of $Re$ and $\Phi$, we define a total reverse momentum of the CRZ, $M$, as

$$M = \int_0^H \int_0^R (2\pi\rho v r)\,dr\,dh, \tag{1}$$

where $\rho$ is the density of the burned gas, $v$ is the axial velocity component, $H$ is the maximum height of the CRZ, and $R$ is the radius of the CRZ boundary at a given height $h$. Here, the boundary of the CRZ is identified as the zero-axial velocity contour.

Fig. 7 plots the variations of $M$ as a function of $Re$ or $\Phi$, for various cases. In general, the variation trends of $M$ are consistent with the qualitative observations obtained by eyeballing the CRZ size in Figs. 5 and 6. We can readily observe that $M$ decreases and approaches zero with increasing $\Phi$ in Fig. 7(a) but increases from zero with increasing $Re$ in Fig. 7(b). As $M = 0$ corresponds well to the cases without CRZ, it serves as an indicator to quantitatively decide the emergence of CRZ.

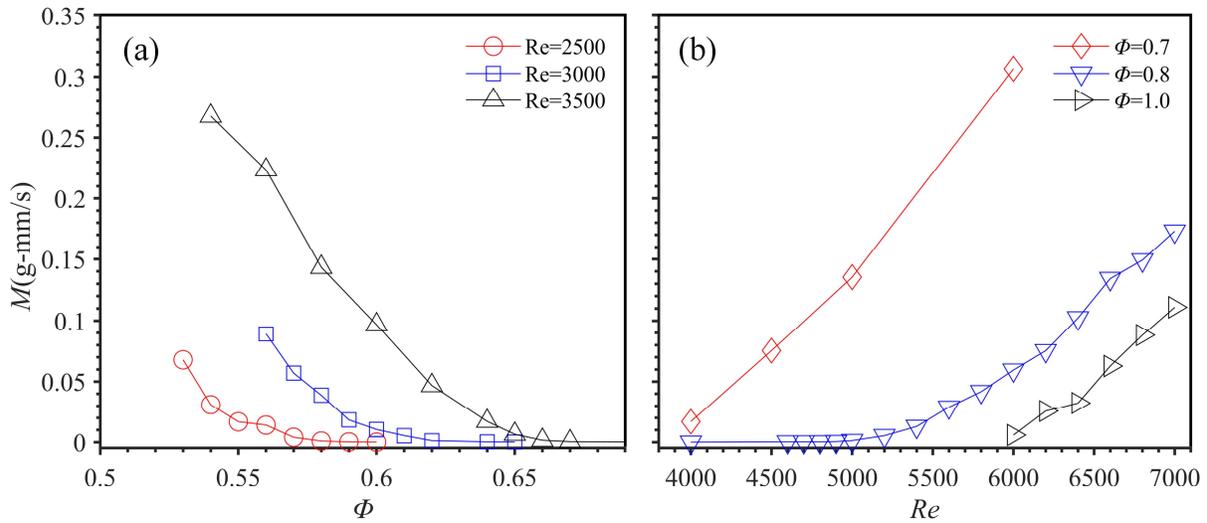

FIG. 7. Dependencies of the CRZ reverse momentum $M$ on $\Phi$ and $Re$.

Swirl intensity is a crucial factor influencing CRZ formation by means of inducing a vortex breakdown (VB) bubble [28,29], and the CRZ size increases with swirl intensity [30]. The relevance of swirl intensity in affecting CRZ emergence is considered for the present work. The swirl intensity is quantified using the swirl number ($S_N$), defined as the ratio between the axial flux of tangential momentum, $G_\theta$, and the axial momentum flux, $G_x$, in the form [31],

$$S_N = \frac{G_\theta}{(D/2)G_x} = \frac{2\pi\rho \int_0^\infty V_x V_\theta r^2 dr}{\pi\rho D \int_0^\infty (V_x^2 - \frac{V_\theta^2}{2})r dr} , \tag{2}$$

where $\rho$ is the gas density, $V_x$ the axial velocity, and $V_\theta$ the tangential velocity. Based on the transverse velocity field measured at 2 mm above the nozzle exit, we calculated $S_N$ for the cases in Fig. 4. We find that for different cases $S_N$ falls into a small range, 0.36~0.38, with a maximum uncertainty of 3.46%. The $S_N$ values also accord with the above observation of the near-constant flame angles for different cases, which can be explained by the strong correlation between the swirl intensity and the flame angle at the base [32,33]. This near-constant $S_N$ is not surprising since only one swirler was adopted in this experiment. This means that in the present study the CRZ formation and the change in the CRZ behaviors, including the size and shape, do not result from the swirling effect. However, considering that swirling influences the CRZ through inducing VB, the underlying implication is profound—the CRZ emergence and characteristics are not merely determined by VB.

## C. Relationships between CRZ and its surrounding ISL

Observing the vorticity field in Figs. 5 and 6, we further note an interesting correlation between the size of the CRZ and the intensity of its surrounding ISL. Specifically, for a smaller-$\Phi$ case in Fig. 5 or a higher-$Re$ case in Fig. 6, the vorticity is higher with a more concentrated distribution, and the ISL is more capable of penetrating downstream. This suggests that enhanced ISL could promote the emergence or increase the size of CRZ as a result of the coupling interaction between the two structures. Such coupling effect was also evident in Stöhr *et al.*'s experiment [34], where the precessing vortex core (PVC) stemmed from the ISL interacts with the CRZ and periodically changes the flame stagnation point. They also found that the OSL has no direct effect on the CRZ because they are separated from each other by the ISL. Zhang *et al.* [35] found that the coupled movements of CRZ and PVC cause the transition of the CRZ shape between asymmetric and symmetric states. In this sense, the change in the CRZ characteristics in the present study is another outcome of the interaction between CRZ and ISL.

From the vortex-dynamics point of view, the CRZ grows as the vorticity in the ISL induces velocity opposite to the mainstream [13,36]. Here, the local strength (or intensity), $\gamma(y)$, of ISL is calculated as the line integration of vorticity across the shear layer in the form, $\gamma(y) = \int_0^\delta \omega dr$, where $\omega$ is the vorticity and $\delta(y)$ is the local width of ISL. The relationship of $|\gamma|$ vs. $y$ for the representative cases in Fig. 5(b) and Fig. 6(a) are plotted in Fig. 8(a) and 8(b), respectively. We can observe that $|\gamma|(0)$ at the nozzle exit is almost a constant for cases of the same $Re$ in Fig. 8(a), whereas it increases with increasing $Re$ for cases of constant $\Phi$ in Fig. 8(b). This means that the initial ISL strength is dictated by the inertia of its upstream swirling jet. However, as the ISL evolves downstream and interacts with the V-shaped flame front, the ISL intensity $|\gamma|$ first increases and then decreases after reaching a maximum value, $|\gamma|_{max}$, for all cases in Fig. 8.

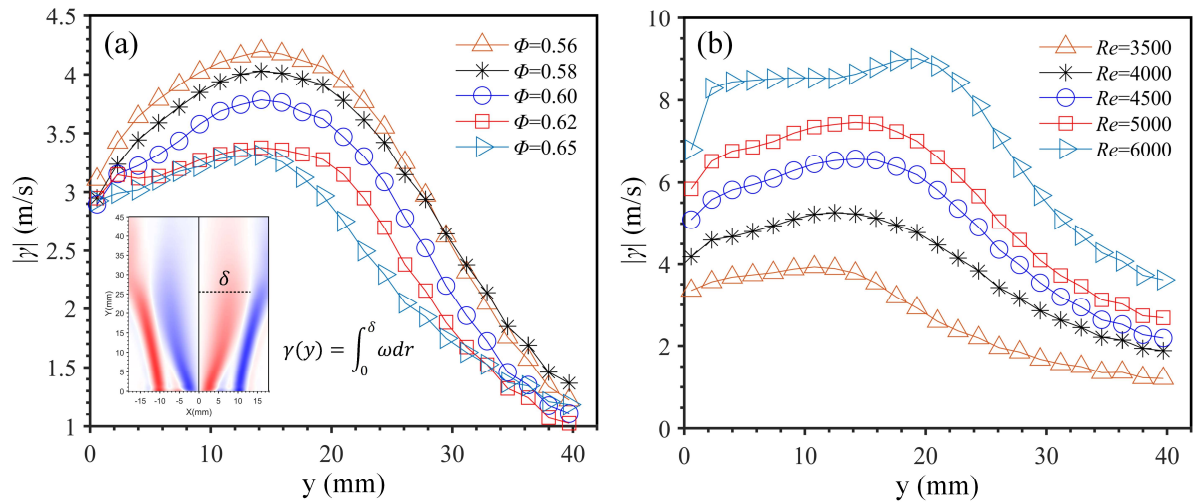

FIG. 8. Variations of the ISL intensity along the $y$ axis for cases with (a) various equivalence ratios at $Re = 3000$, and (b) various Reynolds numbers at $\Phi=0.7$. The subgraph inserted in (a) illustrates the definition of ISL intensity.

The mechanism influencing the variation of ISL intensity may be understood in terms of the vorticity transport equation [37],

$$\frac{D\boldsymbol{\omega}}{Dt} = \underbrace{(\boldsymbol{\omega}\cdot\boldsymbol{\nabla})\boldsymbol{u}}_{stretching/tilting} - \underbrace{\boldsymbol{\omega}(\boldsymbol{\nabla}\cdot\boldsymbol{u})}_{therm\ dilatation} + \underbrace{\frac{1}{\rho^2}(\boldsymbol{\nabla}\rho\times\boldsymbol{\nabla}p)}_{baroclinic\ torque} + \underbrace{\frac{\rho_A}{\rho^2}(\boldsymbol{\nabla}\rho\times\boldsymbol{g})}_{buoyancy} + \underbrace{\nu\boldsymbol{\nabla}^2\boldsymbol{\omega}}_{diffusion}, \quad (3)$$

where $\boldsymbol{u}$ and $\boldsymbol{\omega}$ are the velocity and vorticity vectors, $\rho$ the local density, $p$ the pressure, $\boldsymbol{g}$ the gravitational acceleration vector, $\nu$ the kinematic viscosity, and $\rho_A$ the ambient air density. The terms on the right-hand side of the equation account for the rate of change of vorticity due to stretching/tilting, thermal dilatation, baroclinic torque, buoyancy, and diffusion, respectively. In the case of the ISL, where vorticity is in the azimuthal direction, the azimuthal component of Eq. (3) needs to take the integral form,

$$\int_0^\delta \frac{D\omega}{Dt}dr = \frac{D\gamma}{dt}$$

$$= \int_0^\delta \left[(\boldsymbol{\omega}\cdot\boldsymbol{\nabla})\boldsymbol{u} - \boldsymbol{\omega}(\boldsymbol{\nabla}\cdot\boldsymbol{u}) + \frac{1}{\rho^2}(\boldsymbol{\nabla}\rho\times\boldsymbol{\nabla}p) + \frac{\rho_A}{\rho^2}(\boldsymbol{\nabla}\rho\times\boldsymbol{g}) + \nu\boldsymbol{\nabla}^2\boldsymbol{\omega}\right]_\theta dr. \quad (4)$$

For a qualitative understanding, the order-of-magnitudes of the different terms in Eq. (4) are estimated (see the APPENDIX for details). We find that vortex stretching and baroclinic torque contribute to increasing ISL intensity, whereas thermal dilatation, buoyancy, and vorticity diffusion reduce the ISL intensity. In addition, the buoyancy and diffusion terms are significantly smaller than the thermal dilatation term, implying thermal dilatation is the primary cause of the downstream decay of ISL intensity. The roles of vortex stretching, baroclinic torque, and thermal dilatation in tuning ISL intensity presented here are also consistent with Kiesewetter *et al.*'s [38] numerical research. However, the increase of $|\gamma|$ in the near field could result from a combined effect of vortex stretching and baroclinic torque, which outweighs the thermal dilatation. This happens as the swirling jet near the burner exit has a stronger axial vorticity than the far field, which induces a more significant stretching effect, and a higher negative pressure gradient to yield a larger baroclinic torque. These effects have also been corroborated by our recent numerical data [39].

To further reveal the relationships between the CRZ characteristics and the ISL intensity, $|\gamma|_{max}$ for the same cases in Fig. 7 are plotted in Fig. 9(a) and 9(b), with the black-dashed line separating the cases with and without CRZ. We can observe that for a fixed $Re$ in Fig. 9(a) the ISL intensity decreases with increasing $\Phi$, while it increases with increasing $Re$ for a constant $\Phi$ in Fig .9(b). These trends of $|\gamma|_{max}$ variation agree with the dependences of the CRZ size on $Re$ and $\Phi$ observed from Figs. 5-7, further verifying the correlation between the ISL intensity and the CRZ size. The results seem to demonstrate that the emergence of CRZ is promoted by enhancing $\gamma_{max}$ when either $Re$ or $\Phi$ is fixed. However, the critical $|\gamma|_{max}$ for a constant $Re$ increases with $Re$ and the critical $|\gamma|_{max}$ for a constant $\Phi$ increases with $\Phi$, implying that the emergence of the CRZ could be influenced by effects other than the local ISL intensity.

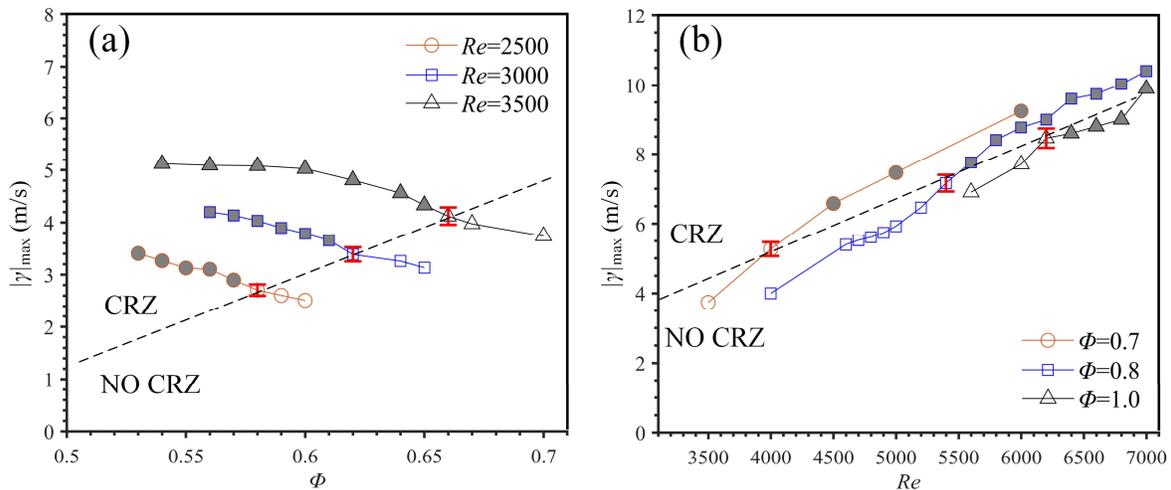

FIG. 9 (a) and (b) plot the max local ISL intensity $|\gamma|_{max}$ as functions of $\Phi$ and $Re$, respectively, for different flames, with the dashed lines separating the cases with and without CRZ. The error bars give the uncertainties in estimating $|\gamma|_{max}$ of the critical cases.

## IV. MECHANISM GOVERNING CRZ EMERGENCE

### A. Dimensional analysis

The above has demonstrated that the ISL significantly influences the CRZ characteristics of the present V-shaped premixed swirling flames. Following this understanding, we present a simple flow model in Fig. 10, based on which dimensional analysis is performed next to explore the mechanism determining CRZ emergence.

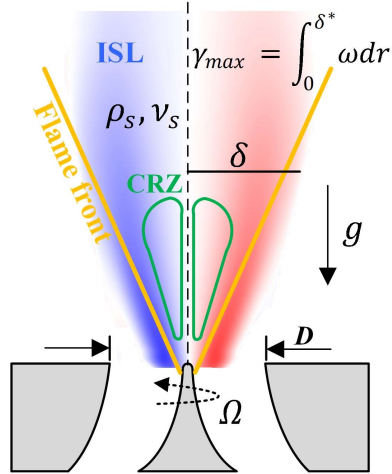

FIG. 10. A simple flow model for a swirling flame with ISL and CRZ.

Let $E$ be a dimensionless quantity distinguishing the cases with and without CRZ, then we can express $E$ in the following relationship,

$$E = F(|\gamma|_{max}, \Omega, D, \rho_s, \nu_s, g), \tag{5}$$

where $\Omega$ is the characteristic angular velocity at the nozzle exit, with the exit diameter $D$ being the characteristic length. The flame in this analysis is treated in the sense of a reacting flow, with its fluid-dynamics effect reflected by the changes in the ISL intensity, $\gamma_{max}$, and the fluid properties, $\rho_s$ and $\nu_s$, associated with the ISL. The flow outside of the ISL is not explicitly considered in this analysis as its effect on the ISL has been accounted for by $|\gamma|_{max}$. Applying the Buckingham Pi theorem [40], we obtain

$$E = F_1\left(\frac{\Omega D}{|\gamma|_{max}}, \frac{|\gamma|_{max}}{\sqrt{gD}}, \frac{|\gamma|_{max}D}{\nu_s}\right) = F_1\left(S, Fr, \frac{|\gamma|_{max}D}{\nu_s}\right), \tag{6}$$

where $S = \frac{\Omega D}{|\gamma|_{max}}$ and $Fr = \frac{|\gamma|_{max}}{\sqrt{gD}}$. $S$ is a characteristic swirl number, which can be considered a constant according to the experimental result in the previous section. $Fr$ for the present study is on the order of $O(10)$, meaning buoyancy plays a negligible role, and thereby $Fr$ can be dropped out. Note that $\frac{|\gamma|_{max}D}{\nu_s}$ has the functionality of Reynolds number, so we can simplify Eq. (6) as

$$E = F_2(Re_s), \tag{7}$$

where

$$Re_s = \frac{|\gamma|_{max}D}{\nu_s}. \tag{8}$$

Eq. (7) clearly shows that $Re_s$ is the governing parameter for the CRZ emergence. We hypothesize that the underlying physics of $Re_s$ could resemble that of the Reynolds number for the classical flow around a cylinder, which is also correlated to the strength of the shear layer in the cylinder wake. Specifically, in the Stokes regime with small Reynolds numbers, the flow passing the cylinder is fully laminar with no separation or recirculation. As the Reynolds number increases to a critical value of ~5 [41,42], flow separation occurs, and a recirculation bubble appears behind the cylinder; further increasing the Reynolds numbers would cause the recirculation bubble to grow larger [43,44]. These trends are in good agreement with those of CRZ emergence and size growth under increasing $\gamma_{max}$ in the present study.

## B. The $Re_s$ criterion

We next explore the validity of $Re_s$ as a criterion for determining the CRZ emergence. To this end, the main challenge of calculating Eq. (8) lies in estimating the kinematic viscosity $\nu_s$ within the flame, which is a function of the local gas temperature $T$ [45]. Experimentally, it requires the application of non-intrusive temperature measurement, which is currently unavailable.

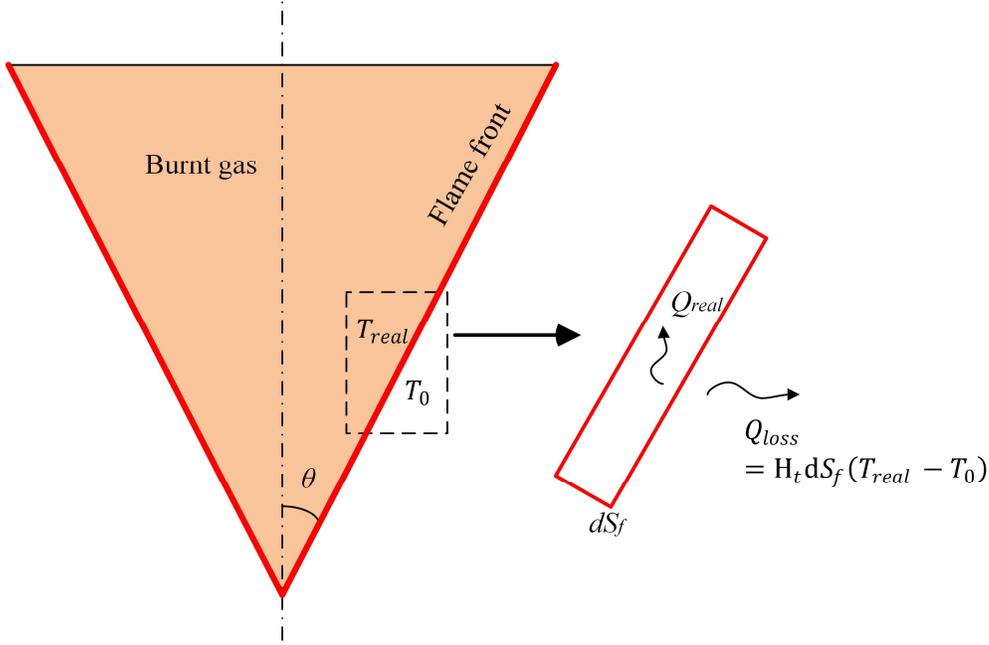

FIG. 11. A simple convective heat transfer model for the V-shaped unconfined flame front.

To estimate the temperature of the burned gas after the flame front, we propose a simple convective heat transfer model, based on balancing the heat release discrepancy at the flame front and the heat loss to the ambient air, as illustrated in Fig. 11. For ideal adiabatic combustion, the heat release $Q_{ad}$ from a flame surface element $dS_f$ can be calculated as

$$Q_{ad} = c_p \bar{m} dS_f (T_{ad} - T_0), \qquad (9)$$

where $T_{ad}$ is the adiabatic flame temperature, $T_0$ the temperature of the unburnt reactants, $c_p$ the effective constant-pressure specific heat, and $\bar{m}$ the consumption rate of reactants at the flame surface. However, owing to heat transfer in reality, the temperature $T_{real}$ of the burnt gas is lower than $T_{ad}$, and has a net heat release of

$$Q_{real} = c_p \bar{m} dS_f (T_{real} - T_0). \tag{10}$$

For the present methane/air flame with a relatively small size, we may neglect the thermal radiation and attribute the heat loss entirely to convection. Thus, the heat loss $Q_{loss}$ from the same flame surface element $dS_f$ takes the form,

$$Q_{loss} = H_t dS_f (T_{real} - T_0), \tag{11}$$

where $H_t$ is the effective convective heat transfer coefficient, and the ambient air also has a temperature of $T_0$. Based on energy conservation, the difference between the adiabatic and the actual heat release rates should result from the convective heat loss, expressed as $Q_{ad} - Q_{real} = Q_{loss}$, which can be combined with Eqs. (9), (10), and (11) to yield

$$c_p \bar{m}(T_{ad} - T_{real}) = H_t(T_{real} - T_0), \tag{12}$$

where $dS_f$ cancels on both sides. Recall the definition of the laminar flame speed $S_L$ for premixed flame [46], $S_L = \frac{\bar{m}}{\rho_u}$, where $\rho_u$ is the density of the unburnt reactants, Eq. (12) can be further derived as

$$\Delta T = T_{ad} - T_{real} = \left(\frac{H_t}{H_t + c_p \rho_u S_L(\Phi)}\right)(T_{ad}(\Phi) - T_0), \tag{13}$$

where we assume $H_t$, $c_p$, and $\rho_u$ to be known constants, and $T_{ad}$ and $S_L$ to be functions of the equivalence ratio $\Phi$. Now, we can use Eq. (13) to estimate $T_{real}$ based on the data of $T_{ad}(\Phi)$ and $S_L(\Phi)$ for methane/air flame [47].

To give a first approximation for Eq. (13), $\rho_u$ takes the density of air at 1atm and 300K, and $c_p$ takes the corresponding value of air at 1atm and 1100K ($\approx 0.5(T_0+T_{ad})$ at an equivalence ratio of 0.75 for methane/air flame). The estimation of $H_t$ is more challenging since the actual flame temperature or the burned gas was not measured in the present work. Assuming flames of similar configuration and flow regime have comparable heat transfer characteristics, we can take advantage of the temperature data of two different V-shaped swirling flames (also fueled by premixed methane/air) reported in the literature [48,49] to compute their effective heat transfer coefficients with Eq. (13); the relevant parameters are listed in Table I, where $T_{exp}$ is the average temperature of the CRZ from measurement. Note that, although the two flames were generated from different combustors and of distinct operating conditions ($Re$, $\Phi$, $S_N$), the resulting $H_t$ ends up to be relatively close. Therefore, we consider the average $H_t$ of these two flames, 0.037 $kW/(m^2K)$, to be a reasonable approximation for the heat transfer coefficient in the simplified model of Eq. (13), as long as the flame is within the same regime. The resultant estimation of the normalized $\Delta T_n$, defined as $\Delta T/\Delta T(\Phi = 1)$, is plotted in Fig. 12, showing an interesting trend that the temperature gap between $T_{ad}$ and $T_{real}$ decreases with increasing $\Phi$ for lean combustion. This is somewhat counter-intuitive as one might expect $\Delta T$ to increase because heat loss increases at higher $T_{ad}$ as $\Phi$ approaches 1. However, since $\Phi$ has a larger influence on $S_L(\Phi)$ than on $T_{ad}(\Phi)$, the increased heat loss with increasing $\Phi$ is overwhelmed by the more enhanced burning rate associated with the elevated flame speed, which according to Eq. (13) would yield a smaller $\Delta T$.

Table I. Parameters of two different flames [48,49] for the estimation of $H_t$.

| No. | $Re$ | $\Phi$ | $S_N$ | $T_0(K)$ | $T_{ad}(K)$ | $T_{exp}(K)$ | $H_t(kW/m^2K)$ |
|---|---|---|---|---|---|---|---|
| 1 | 10000 | 0.83 | 0.75 | 300 | 2040 | 1880 | 0.0381 |
| 2 | 4900 | 0.8 | 0.6 | 300 | 1997 | 1817 | 0.0357 |

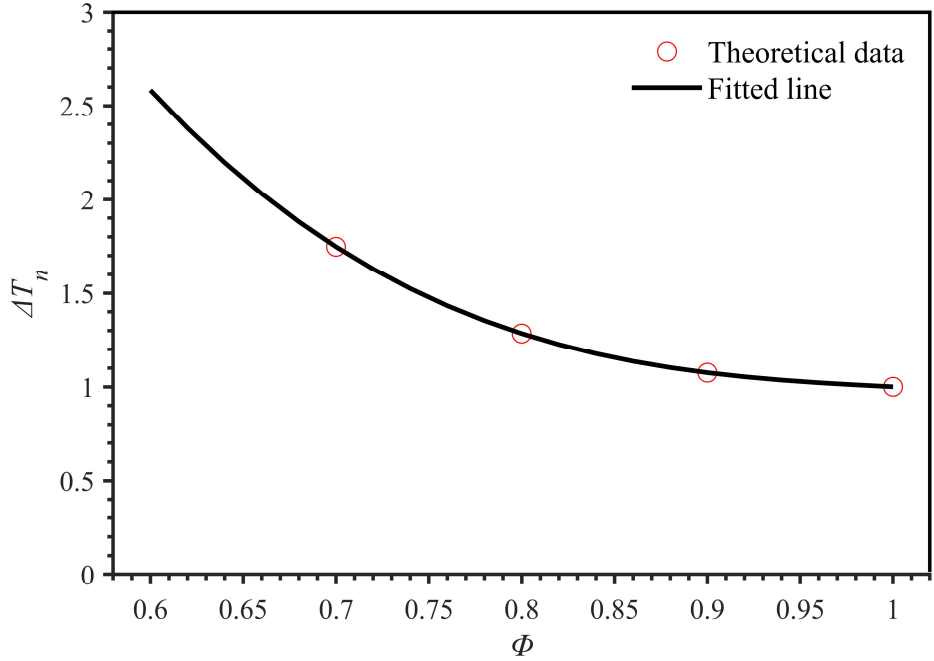

FIG. 12. Variation of $\Delta T_n$ with $\Phi$ calculated from Eq. (13), based on the methane/air flame data in the literature [47].

With $T_{real}$ obtained from Eq. (13), we can further estimate the viscosity $\nu_s$ using a similar method to Oijen [45]. Then, the $Re_s$ criterion for CRZ emergence is examined in Fig. 13. We see that a single linear boundary line through the origin exists between the regions with and without the CRZ; and the critical $Re_s$ is determined to be 424 ± 11 based on the slope of this fitted line, justifying the effectiveness of the proposed $Re_s$ criterion in determining the CRZ emergence in the present V-shaped premixed swirling flames. Note the actual uncertainty with $Re_s$ could be larger since the accuracy of the termperature model in Eq. (13) was not measured.

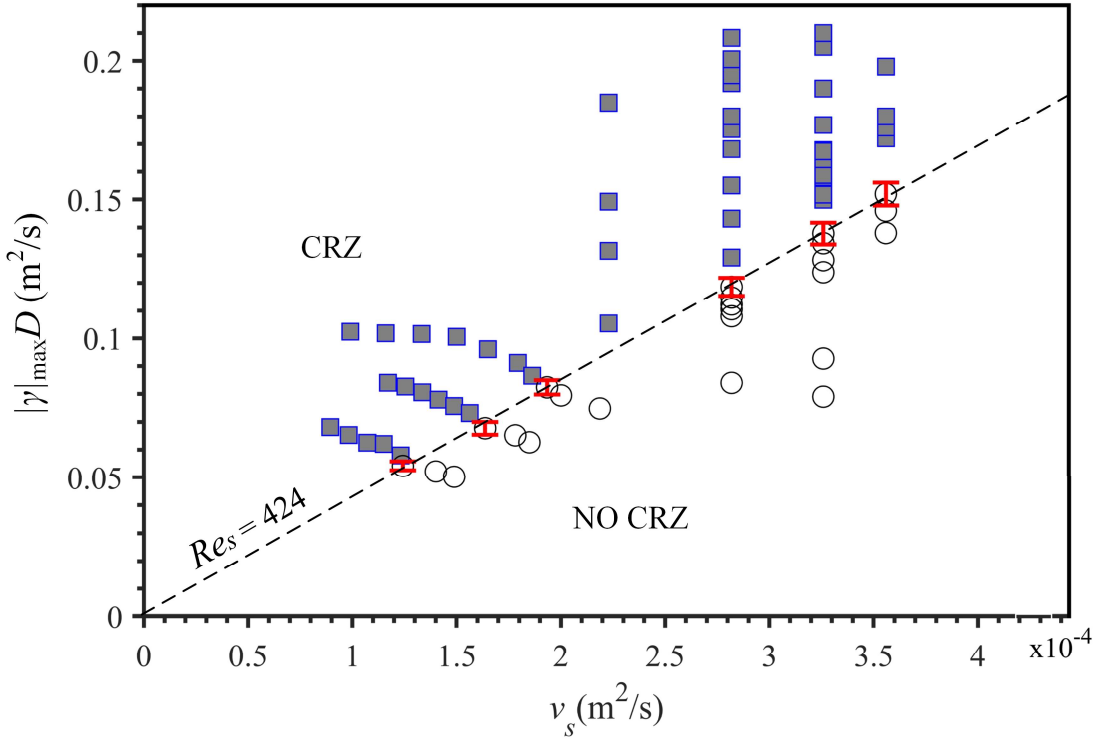

FIG. 13. The relation of $|\gamma|_{\max}D$ vs. $\nu_s$, with $\nu_s$ evaluated based on the flame temperature calculated from Eq. (13). The boundary between CRZ and No CRZ collapses to a single line for all test cases, yielding a critical $Re_s$ of $424 \pm 11$. Error bars are shown for the critical cases near the boundary.

Before closing, it should be mentioned that, while the present V-shaped flames belong to the low-swirl regime, the conclusions regarding the influences of the ISL at the nozzle exit on CRZ may be extended to the high-swirl cases with some modifications, as the underlying mechanism is similar. Specifically, an enhanced swirl intensity is likely to cause a larger flame angle and promote the formation tendency of the CRZ, which could be reflected in a different critical $Re_s$. These will be addressed in future works.

## V. CONCLUDING REMARKS

In this paper, the emergence of the central recirculation zone (CRZ) in a V-shaped premixed swirling flame was experimentally investigated using a simultaneous measurement of PIV and CH* chemiluminescence. By tuning the Reynolds number $Re$ and equivalence ratio $\Phi$ of the methane-air premixture, it was found that both increasing the $Re$ and decreasing the $\Phi$ could facilitate the emergence of the CRZ. Further analysis demonstrates that the CRZ characteristics and its emergence are strongly influenced by the inner shear layer (ISL) surrounding the CRZ.

To reveal the exact mechanism determining the CRZ emergence, dimensional analysis was performed to yield a non-dimensional critical parameter, $Re_s = |\gamma|_{max} D / \nu_s$, defined based on the max ISL intensity ($|\gamma|_{max}$), the exit diameter ($D$), and the kinematic viscosity ($\nu_s$) of the burnt gas. The underlying physics of $Re_s$ resembles that of the Reynolds number for the classical flow around a cylinder, for which a recirculation bubble appears as $Re$ increases to a critical value. Further incorporating a simplified model to estimate the ISL temperature and viscosity, we plotted the $|\gamma|_{max} D - \nu_s$ relationship and obtained a single boundary line separating the cases with and without CRZ. This points to the existence of a common critical $Re_s$, justifying its effectiveness for lean-premixed V-shaped swirling flames of various operating conditions.

Last, the significance of the present work is emphasized. In general, the CRZ emergence in swirling flames can be explained from the point of vortex breakdown because it dictates the upstream onset of CRZ. The current result brings to our attention the role of ISL intensity. Specifically, flame heating could attenuate the ISL and further suppress the CRZ downstream.

## ACKNOWLEDGMENTS

This work was supported partly by the National Natural Science Foundation of China (Grant No. 52006139, 92041001, and 12072194) and partly by the Shanghai Sailing Program (Grant No. 20YF1420600).

# APPENDIX

This appendix provides an order-of-magnitude estimation of the vorticity variations in the ISL contributed by the different terms in Eq. (4). We write Eq. (4) in cylindrical coordinate $(r, \theta, z)$:

$$\frac{D\gamma}{Dt} = \underbrace{\int_0^\delta \left(\omega_r \frac{\partial u_\theta}{\partial r} + \omega_z \frac{\partial u_\theta}{\partial z} + \frac{u_r \omega_\theta}{r}\right) dr}_{stretching/tilting} - \underbrace{\int_0^\delta \omega_\theta \left[\frac{1}{r}\frac{\partial(r u_r)}{\partial r} + \frac{\partial u_z}{\partial z}\right] dr}_{thermal\ dilatation} - \underbrace{\int_0^\delta \frac{\rho_A}{\rho^2}\left(\frac{\partial \rho}{\partial r} g_z\right) dr}_{buoyancy}$$

$$+ \underbrace{\int_0^\delta \frac{1}{\rho^2}\left(\frac{\partial \rho}{\partial z}\frac{\partial p}{\partial r} - \frac{\partial \rho}{\partial r}\frac{\partial p}{\partial z}\right) dr}_{baroclinic\ torque} + \underbrace{\int_0^\delta \nu \left(\frac{\partial^2 \omega_\theta}{\partial r^2} + \frac{1}{r}\frac{\partial \omega_\theta}{\partial r} + \frac{\partial^2 \omega_\theta}{\partial z^2} - \frac{\omega_\theta}{r^2}\right) dr}_{diffusion}, \qquad (A1)$$

where all the $\partial\theta$ derivatives are dropped out under the axisymmetric condition, and the vorticity components can be calculated as

$$\omega_r = -\frac{\partial u_\theta}{\partial z}, \qquad (A2)$$

$$\omega_\theta = \frac{\partial u_r}{\partial z} - \frac{\partial u_z}{\partial r}, \qquad (A3)$$

$$\omega_z = \frac{1}{r}\frac{\partial(r u_\theta)}{\partial r} = \frac{u_\theta}{r} + \frac{\partial u_\theta}{\partial r}. \qquad (A4)$$

Note that, in the cylindrical coordinate, $\omega_\theta$ has an opposite sign to that in the Cartesian coordinate owing to the inward-pointing azimuthal direction, so $\gamma$ obtained here also has an opposite sign. In this work, we estimate the terms in Eq. (A1) based on the flow field data at 2 mm above the nozzle exit, where the velocity and azimuthal vorticity profiles for a representative case with $\Phi = 0.6$ and $Re = 3000$ are plotted in Fig. A1. The average magnitudes of $u_z$, $u_r$, $u_\theta$, and $\omega_\theta$ within the ISL are on the orders of $10^0$, $10^{-1}$, $10^0$, and $10^2$, respectively, based on the SI units. Fig. A1 shows that within the ISL $u_r$, $u_z$, and $u_\theta$ increases monotonically with $r$, while $\omega_\theta$ first decreases and then increases; the profiles for other cases have similar trends. For first-order approximations, the velocities and azimuthal vorticity in the ISL are assumed to depend linearly on $r$:

$$u_z = Ar, \qquad (A5)$$

$$u_r = Br, \qquad (A6)$$

$$u_\theta = Cr, \qquad (A7)$$

$$\omega_\theta = \begin{cases} -ar, & \text{if } 0 \leq r \leq \delta_m \\ -b + cr, & \text{if } \delta_m < r \leq \delta' \end{cases} \qquad (A8)$$

where $A, B, C, a, b$, and $c$ are on the orders of $10^3, 10^2, 10^2, 10^5, 10^3$, and $10^5$, respectively. $\delta_m$ represents the radial position where $\omega_\theta$ reaches its minimum and is of the same order as $\delta$. Next, we evaluate the individual terms in Eq. (A1).

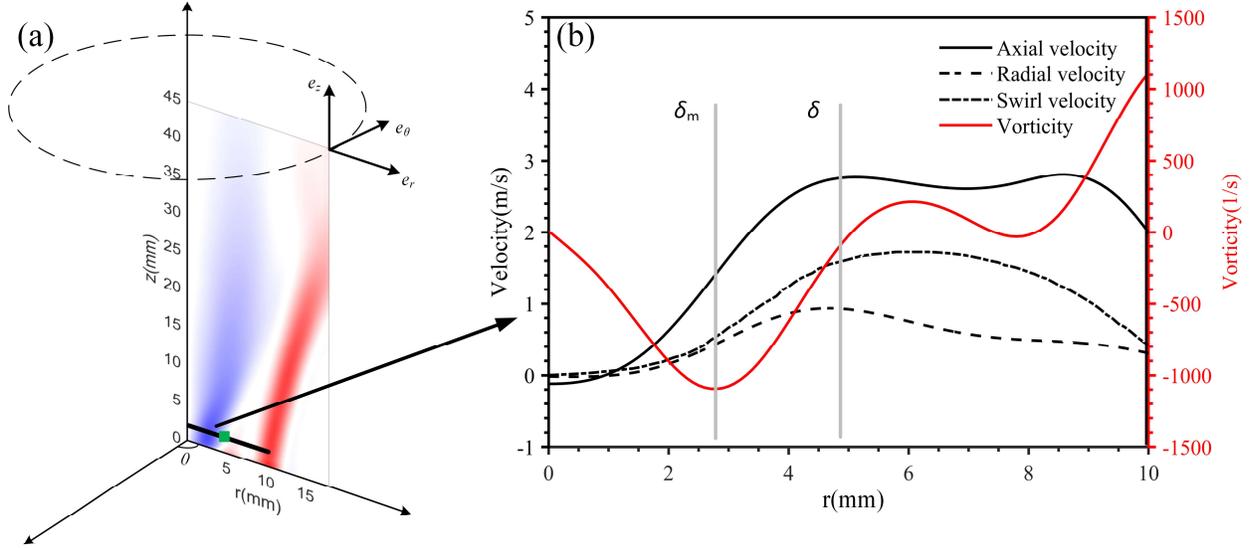

FIG. A1. (a) The azimuthal vorticity distribution in the cylindrical coordinate for the case with $\Phi = 0.6$ at $Re = 3000$. The black line represents the cross-section 2 mm above the nozzle exit, and the green dot indicates the position of the ISL boundary. (b) The velocity and azimuthal vorticity profiles of the cross-section in (a).

(1) ***The stretching/tilting term***. Applying Eqs. (A2-A4), the ISL intensity variation, $\dot{\gamma}_s$, due to vortex stretching and tilting can be expressed as

$$\dot{\gamma}_s = \int_0^\delta \left( \omega_r \frac{\partial u_\theta}{\partial r} + \omega_z \frac{\partial u_\theta}{\partial z} + \frac{u_r \omega_\theta}{r} \right) dr = \int_0^\delta \left( \frac{u_\theta}{r} \frac{\partial u_\theta}{\partial z} + \frac{u_r \omega_\theta}{r} \right) dr. \quad (A9)$$

Substituting Eqs. (A6-A8) into Eq. (A9), $\dot{\gamma}_s$ can be derived as

$$\dot{\gamma}_s = \int_0^\delta C \frac{\partial C}{\partial z} r \, dr - \int_0^{\delta_m} Bar \, dr + \int_{\delta_m}^\delta (Bcr - Bb) dr$$

$$= C \frac{\partial C}{\partial z} \frac{\delta^2}{2} - Ba \frac{\delta_m^2}{2} - Bb(\delta - \delta_m) + Bc \frac{(\delta^2 - \delta_m^2)}{2}, \quad (A10)$$

where both $C$ and $\delta$ depend on $z$. The second to fourth terms in Eq. (A10) are on the orders of $-10^1$, $-10^2$, and $10^1$, respectively. To determine $\partial C/\partial z$, we assume that in the vicinity of the jet exit the total circulation of ISL in the horizontal plane remains an invariant of $z$, which means

$$\frac{\partial (u_\theta(r=\delta) 2\pi\delta)}{\partial z} = \frac{\partial (2\pi C \delta^2)}{\partial z} = 0. \quad (A11)$$

Solving Eq. (A11) gives $\partial C/\partial z = -2C \partial \delta/(\delta \partial z)$. Further note that $\partial \delta/\partial z$ is related to the spread angle of the flame and should be order 1 for V-shaped flames, then we obtain $\partial C/\partial z \sim -2C/\delta$ and its magnitude is on the order of $-10^5$, yielding an order of $-10^1$ for the first term in Eq. (A10). Finally, the overall $\dot{\gamma}_s$ takes the order of $-10^2$.

(2) ***The thermal dilatation term***. The ISL intensity variation, $\dot{\gamma}_t$, due to thermal dilatation can be calculated as

$$\dot{\gamma}_t = -\int_0^\delta \omega_\theta \left[ \frac{\partial u_r}{\partial r} + \frac{u_r}{r} + \frac{\partial u_z}{\partial z} \right] dr. \quad (A12)$$

Substituting Eqs. (A6-A8) into Eq. (A12), yields

$$\dot{\gamma}_t = \int_0^{\delta_m} ar \left[\frac{\partial (Br)}{\partial r} + \frac{Br}{r} + \frac{\partial u_z}{\partial z}\right] - \int_{\delta_m}^{\delta} (cr-b)\left[\frac{\partial (Br)}{\partial r} + \frac{Br}{r} + \frac{\partial u_z}{\partial z}\right] dr$$

$$= \int_0^{\delta_m} ar \left[2B + \frac{\partial u_z}{\partial z}\right] dr - \int_{\delta_m}^{\delta} (cr-b)\left[2B + \frac{\partial u_z}{\partial z}\right] dr. \tag{A13}$$

The variations of $u_z$ with $z$ near 2 mm above the nozzle exit for a representative case with $\Phi = 0.6$ and $Re = 3000$ are ploted in Fig. A2(a). Fig. A2(b) shows that $\partial u_z/\partial z$ first decreases and then increases with $r$ in the ISL, and the minimum is reached at approximately the same radial location $\delta_m$ as that of the minimum $\omega_\theta$. For first-order approximation, we again assume $\partial u_z/\partial z$ in the ISL depends linearly on $r$:

$$\frac{\partial u_z}{\partial z} = \begin{cases} -mr - n, & \text{if } 0 \leq r \leq \delta_m \\ pr - q, & \text{if } \delta_m < r \leq \delta \end{cases}, \tag{A14}$$

where $m, n, p$, and $q$ are on the orders of $10^5, 10^1, 10^5$, and $10^2$, respectively. Substituting Eq. (A14) into Eq. (A13), we obtain

$$\dot{\gamma}_t = a\left[(2B-n)\frac{\delta_m^2}{2} - \frac{m\delta_m^3}{3}\right] - c\left[(2B-q)\frac{(\delta^2 - \delta_m^2)}{2} + \frac{p(\delta^3 - \delta_m^3)}{3}\right]$$

$$+ b \int_{\delta_m}^{\delta} (2B + pr - q) dr. \tag{A15}$$

The magnitudes of the three terms in Eq. (A15) are on the order of $10^1$, $10^1$ and $10^2$, respectively. So the third term determines the overall order of $\dot{\gamma}_t$, but its sign needs to be further decided. Given that B is about 230 from Fig. A1 and $(pr - b)$ is larger than $-250$ from Fig. A2, we have $(2B + pr - b) > 0$. Thus, $\dot{\gamma}_t$ is on the order of $10^2$.

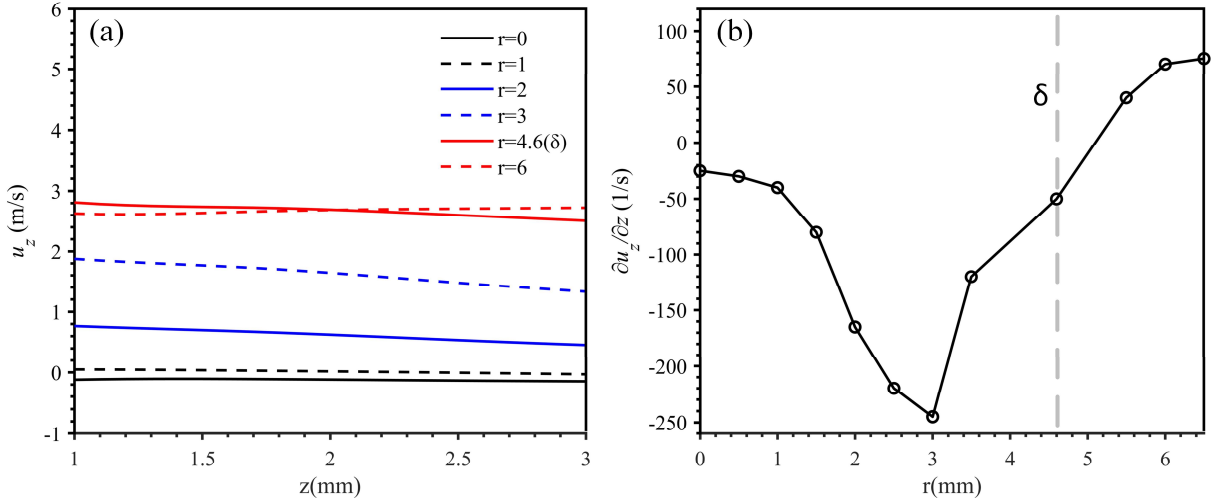

FIG. A2. (a) Variations of $u_z$ with $z$ near 2 mm above the nozzle exit for a representative case with $\Phi = 0.6$ and $Re = 3000$. (b) Variations of $\partial u_z / \partial z$ with $r$ at 2 mm above the nozzle exit for the same case.

(3) **The baroclinic torque term**. For the present flame, it is reasonable to assume that the density gradient exists only in the radial direction, meaning $\partial \rho / \partial z = 0$, so the ISL intensity variation, $\dot{\gamma}_b$, due to baroclinic torque can be reduced as

$$\dot{\gamma}_b = \int_0^\delta \frac{1}{\rho^2}\left(\frac{\partial \rho}{\partial z}\frac{\partial p}{\partial r} - \frac{\partial \rho}{\partial r}\frac{\partial p}{\partial z}\right)dr = -\int_0^\delta \frac{1}{\rho^2}\frac{\partial \rho}{\partial r}\frac{\partial p}{\partial z}dr. \qquad (A16)$$

To simplify the problem, we further apply the flame-sheet approximation so that the density inside or outside the flame can be considered constant while it jumps across the thin flame front. As such, the density can be expressed as a step function in the form,

$$\rho(r) = \begin{cases} \rho_f, & \text{for } r \leq \delta_f \\ \rho_a, & \text{for } r > \delta_f \end{cases}, \qquad (A17)$$

where $\rho_f$ and $\rho_a$ are the gas densities for the hot flame region and cold ambient premixture, respectively, and $\delta_f$ denotes the radial location of the flame front. It follows that Eq. (A16) can be derived as

$$\dot{\gamma}_b = \int_{\rho(\delta_{f-})}^{\rho(\delta_{f+})} \frac{\partial p}{\partial z} d\left(\frac{1}{\rho}\right) = \left(\frac{\partial p}{\partial z}\right)\bigg|_{\delta_f} \left(\frac{1}{\rho_a} - \frac{1}{\rho_f}\right). \quad (A18)$$

To estimate the density-related terms, we note that $\rho_a > \rho_f \sim 1$, so the order of $(1/\rho_a - 1/\rho_f)$ is about $-10^0$. The axial pressure gradient at the flame front can be roughly estimated based on the Bernoulli equation,

$$\frac{\partial}{\partial s}\left(p_1 + \frac{1}{2}\rho_a v_1^2\right) = 0, \quad (A19)$$

where $p_1$ and $v_1$ are the pressure and velocity of a point on a spiral streamline next to but outside the flame front, $s$ is the distance along the streamline. From Fig. A1, the axial velocity component dominates over the other two components, so $\partial/\partial s \sim \partial/\partial z$ and Eq. (A19) can be reformed as

$$\frac{\partial p_1}{\partial z} \sim -\frac{\rho_a}{2}\frac{\partial (v_1^2)}{\partial z} = -\rho_a\left(u_{1z}\frac{\partial u_{1z}}{\partial z} + u_{1r}\frac{\partial u_{1r}}{\partial z} + u_{1\theta}\frac{\partial u_{1\theta}}{\partial z}\right), \quad (A20)$$

where $v_1^2 = u_{1z}^2 + u_{1r}^2 + u_{1\theta}^2$. Based on Figs. A1 and A2, $u_{1z}$, $u_{1r}$, $u_{1\theta}$, and $\partial u_{1z}/\partial z$ are on the orders of $10^0$, $10^{-1}$, $10^0$, and $-10^2$, respectively. Based on Fig. A3, the order of $u_{1r}/\partial z$ is about $-10^2$. Based on the result of Eq. (A11), $\partial u_{1\theta}/\partial z$ has the order of $-10^2$. Therefore, $\partial p_1/\partial z$ is on the order of $10^2$. Finally, $\dot{\gamma}_b$ is on the order of $-10^2$.

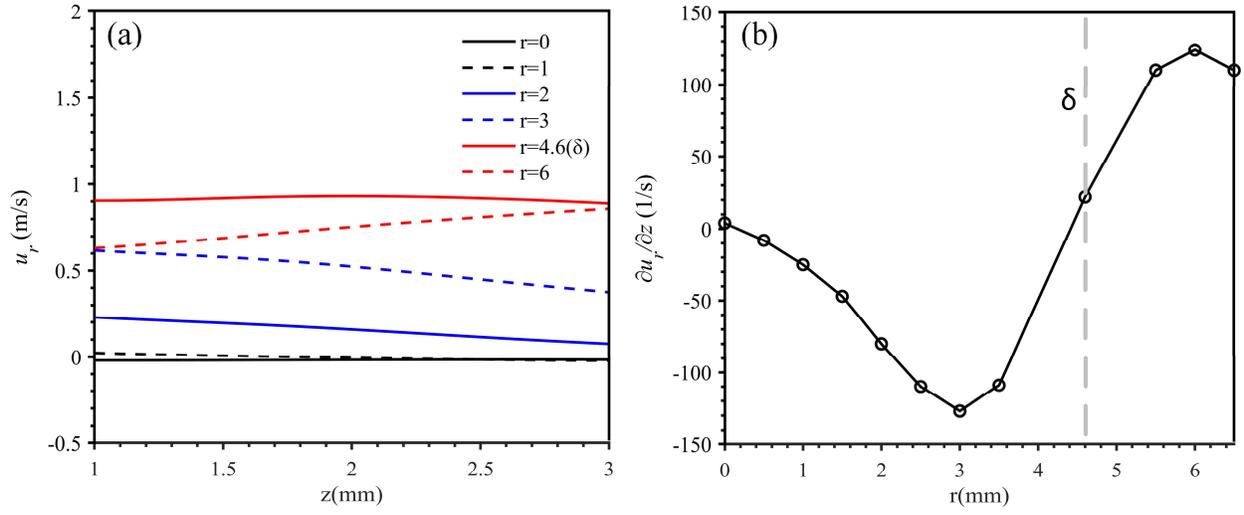

FIG. A3. (a) Variations of $u_r/\partial z$ with $z$ near 2 mm above the nozzle exit for a representative case with $\Phi = 0.6$ and $Re = 3000$. (b) Variations of $\partial u_r/\partial z$ with $r$ at 2 mm above the nozzle exit for the same case.

(4) **The buoyancy term**. For constant gravitational acceleration, $g_z = -9.81$ m/s², and ambient air density $\rho_A$, the ISL intensity variation, $\dot{\gamma}_g$, due to buoyancy can be derived as

$$\dot{\gamma}_g = -\rho_A g_z \int_0^\delta \frac{1}{\rho^2}\frac{\partial \rho}{\partial r} dr = \rho_A g_z \int_0^\delta \frac{\partial(1/\rho)}{\partial r} dr. \tag{A21}$$

Similar to the derivation of Eq. (A16), under the flame-sheet approximation, Eq. (A21) can be simplified as

$$\dot{\gamma}_g = \rho_A g_z \int_{\rho(\delta_{f-})}^{\rho(\delta_{f+})} d\left(\frac{1}{\rho}\right) = \rho_A g_z \left(\frac{1}{\rho_a} - \frac{1}{\rho_f}\right), \tag{A22}$$

which yields an order of $10^1$.

(5) **The diffusion term**. Realizing that the diffusion of vorticity generally occurs in the direction normal to the shear layer, we can assume the diffusion here is primarily in the radial direction, meaning $\partial^2 \omega_\theta / \partial r^2 \gg \partial^2 \omega_\theta / \partial z^2$. Then the ISL intensity variation, $\dot{\gamma}_d$, due to vorticity diffusion can be simplified as

$$\dot{\gamma}_d = \int_0^\delta \nu \left[\frac{\partial^2 \omega_\theta}{\partial r^2} + \frac{\partial(\omega_\theta/r)}{\partial r} + \frac{\partial^2 \omega_\theta}{\partial z^2}\right] dr \sim \int_0^\delta \nu \left[\frac{\partial}{\partial r}\left(\frac{\partial \omega_\theta}{\partial r}\right) + \frac{\partial(\omega_\theta/r)}{\partial r}\right] dr. \quad (A23)$$

For simplicity, the dynamic viscosity $\nu$ is considered a constant of the order $10^{-5}$, so the above equation can be further derived as

$$\dot{\gamma}_d \sim \nu \left(\frac{\partial \omega_\theta}{\partial r}\right)\Big|_0^\delta + \nu \left(\frac{\omega_\theta}{r}\right)\Big|_0^\delta = \nu(c + a), \quad (A28)$$

where Eq. (8) is adopted in estimating the $\partial \omega_\theta / \partial r$ term, and $\omega_\theta$ vanishes at both the centerline ($r = 0$) and the ISL boundary ($r = \delta$). Therefore, the overall $\dot{\gamma}_d$ is on the order of $10^0$.

With the above analysis, the contributions of the stretch/tilting, thermal dilatation, baroclinic, buoyancy, and diffusion terms to the ISL strength variation are estimated to be $-10^2$, $10^2$, $-10^2$, $10^1$, and $10^0$, respectively. Given the negative vorticity within the ISL in the cylindrical coordinate, the vortex stretch/tilting and baroclinic torque serve to strengthen the ISL, whereas the thermal dilatation, buoyancy, and diffusion effects cause the ISL to decay.

---